\documentclass[aps,prx,twocolumn,showpacs,superscriptaddress,footinbib]{revtex4-1} 

\usepackage[utf8]{inputenc}
\usepackage[T1]{fontenc}
\usepackage{lmodern}

\usepackage[english]{babel}
\usepackage{xcolor}
\usepackage{comment}
\usepackage{graphicx}
\usepackage{stackengine}
\usepackage{dcolumn}
\usepackage{bm}
\usepackage{bbm}
\usepackage{amssymb}
\usepackage{amsmath}
\usepackage{bbold}
\usepackage{pstricks}
\usepackage{slashed}
\usepackage{braket}
\usepackage{hyperref}
\usepackage{natbib}
\usepackage{float}
\usepackage{verbatim}
\usepackage{siunitx}
\usepackage{lipsum}
\usepackage{ulem}
\usepackage[left=1.85cm,right=1.85cm,top=1.85cm,bottom=1.85cm]{geometry}

\usepackage{MnSymbol}

\definecolor{mygreen}{rgb}{0,0.5,0}
\definecolor{myblue}{rgb}{0,0,0.75}
\definecolor{mymagenta}{cmyk}{0,1,0,0.12}
\definecolor{mygray}{rgb}{0.5,0.5,0.5}

\newcommand{\Eq}[1]{Eq.~(\ref{#1})}

\newcommand{\Fig}[1]{Fig.~\ref{#1}}

\begin{document}
\title{Rare-Event Quantum Sensing using Logical Qubits}

\author{Robert Ott}
\email{robert.ott@uibk.ac.at}
\affiliation{Institute for Theoretical Physics, University of Innsbruck, Innsbruck, 6020, Austria}\affiliation{Institute for Quantum Optics and Quantum Information of the Austrian Academy of Sciences, Innsbruck, 6020, Austria}

\author{Torsten V. Zache}
\affiliation{Institute for Theoretical Physics, University of Innsbruck, Innsbruck, 6020, Austria}
\affiliation{Institute for Quantum Optics and Quantum Information of the Austrian Academy of Sciences, Innsbruck, 6020, Austria}

\author{Soonwon Choi}
\affiliation{Center for Theoretical Physics—a Leinweber Institute, Massachusetts Institute of Technology, Cambridge, MA 02139, USA}

\author{Adam M. Kaufman}
\affiliation{JILA, University of Colorado and National Institute of Standards and Technology,
and Department of Physics, University of Colorado, Boulder, Colorado 80309, USA}

\author{Hannes Pichler}
\affiliation{Institute for Theoretical Physics, University of Innsbruck, Innsbruck, 6020, Austria}\affiliation{Institute for Quantum Optics and Quantum Information of the Austrian Academy of Sciences, Innsbruck, 6020, Austria}

\begin{abstract}
We present a novel protocol to detect rare signals in a noisy environment using quantum error correction (QEC). The key feature of our protocol is the discrimination between signal and noise through distinct higher-order correlations, realized by the non-linear processing that occurs during syndrome extraction in QEC. In this scheme, QEC has two effects: First, it sacrifices part of the signal $\epsilon$ by recording a reduced, stochastic, logical phase $\phi_L=\mathcal{O}(\epsilon^3)$. Second, it corrects the physical noise and extends the (logical) coherence time for signal acquisition. For rare signals occurring at random times in the presence of local Markovian noise, we explicitly demonstrate an improved sensitivity of our approach over more conventional sensing strategies.
\end{abstract}

\maketitle

\textit{Introduction.---}Quantum metrology~\cite{giovannetti2006quantum,degen2017quantum,pezze2018quantum} is one of the prime applications of quantum technologies~\cite{maze2008nanoscale,hinkley2013atomic,bloom2014optical,abbott2016observation,parker2018measurement,overstreet2022observation,bonus2025ultrasensitive}.
By exploiting entanglement~\cite{nagata2007beating,gross2010nonlinear,leroux2010implementation,ligo2011gravitational,muessel2014scalable,lachance2020entanglement,gilmore2021quantum,eckner2023realizing,bornet2023scalable}, quantum sensors can achieve sensitivities which improve linearly with the number of involved particles. This scaling is known as the 'Heisenberg limit' (HL), and it constitutes a significant improvement over the 'standard quantum limit' (SQL) reachable with uncorrelated probes. However, this conceptual advantage of entangled quantum sensors is typically rapidly lost in practical applications~\cite{crawford2021quantum,aslam2023quantum} due to decoherence~\cite{huelga1997improvement,fujiwara2008fibre,escher2011general,demkowicz2012elusive,chaves2013noisy,demkowicz2014using}.

While the same challenge arises in quantum computing~\cite{bluvstein2024logical,google2025quantum,brock2025quantum,krinner2022realizing,schindler2011experimental,putterman2025hardware}, quantum error correction (QEC)~\cite{terhal2015quantum} provides an efficient framework to protect quantum operations, by encoding and monitoring quantum information in suitably entangled many-particle states. Accordingly, a natural question is whether quantum metrology might also benefit from analogous correction strategies~\cite{dur2014improved,kessler2014quantum,arrad2014increasing,Unden2016,layden2019ancilla}.
Consideration of this question for Markovian noise reveals that, when the signal shares certain features with the noise---more precisely, if "the signal Hamiltonian lies in the span of the Lindblad operators"~\cite{demkowicz2017adaptive,sekatski2017quantum,zhou2018achieving,zhou2021asymptotic}---QEC for quantum metrology is fundamentally limited and therefore not helpful in many physical circumstances.

\begin{figure}[t]
    \flushleft    \includegraphics[width=.96\linewidth]{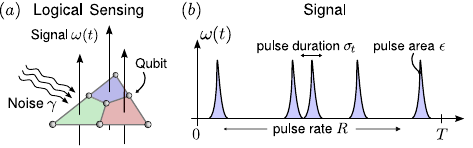}
    \caption{\textit{QEC-enhanced logical sensing scheme.} ($a$) We entangle physical sensors and form quantum codes, for example a 7-qubit Steane code, to improve the sensing performance in the presence of local Markovian decoherence with strength~$\gamma$. ($b$)~We consider the task of sensing a random rare (with known rate $R$) and weak signal $\omega(t)$ in the form of short pulses with area~$\epsilon$ and duration~$\sigma_t$.}
    \label{fig:Overview}
\end{figure}

In this letter, we present a novel type of quantum sensing protocol where QEC can enhance sensor performance for general local Markovian noise [\Fig{fig:Overview}($a$)]. Our protocol applies to pulsed signals that are both \emph{weak}, parametrized by a small-angle rotation~$\epsilon$, and also \emph{rare}, with a known, small, average pulse rate~$R$, see \Fig{fig:Overview}($b$).
For such signals we demonstrate an advantage of our \emph{logical} sensing strategy over standard \emph{physical} sensing strategies when the physical noise rate~$\gamma$ dominates over the signal rate, i.e.~$\gamma \gg R$.
This is possible because the signal, after applying syndrome extraction and error correction, realizes a higher-order logical phase rotation~\footnote{Similar approaches have recently been explored to stochastically implementing small-angle transversal logical gates in quantum codes}\cite{huang2025robust,choi2023fault,ismail2025transversal,Yoshioka2025Transversal}.
The resulting logical signal can be coherently acquired and read-out via logical measurements at the end of each sensing cycle. 
    While the error-correction procedure attenuates the signal strength from $\epsilon$ to $\propto \epsilon^3$, the error correction allows for an extended coherence time of the logical sensors, which more than offsets the cubic reduction in signal. We specifically demonstrate this behavior using both analytical calculations and numerical simulations of our protocol for a $[[7,1,3]]$ Steane code, and discuss generalizations to other codes further below.

\textit{Signal structure and sensing task.---}In this paper, we consider a special type of signal consisting of a sequence of short, randomly spaced pulses, see~\Fig{fig:Overview}($b$).
We model the signal amplitude as $\omega(t) = \sum_i  f(t-t_i)$, where each pulse has a shape $f$ with pulse area $\epsilon = \int\mathrm{d}t\, f(t)$, and duration $\sigma_t$ much shorter than all other relevant time scales, such that it is effectively described by a Dirac-delta function $f(t)=\epsilon \delta(t)$. To be concrete, we assume the pulses to arrive at random, unknown times $t_i$, distributed according to a Poisson process with a known mean pulse rate $R$. In this setting, we consider the task of determining $\epsilon$.

For this we consider a setup of $n$ physical sensors that are exposed to the signal for a total sensing time $T$. We model the sensors as two-level systems, $\{\ket{0}_j,\ket{1}_j\}$, $j=1,..,n$, which couple homogeneously to the signal via the time-dependent Hamiltonian
\begin{align}
\label{eq:Hamiltonian}
    H(t) = \omega(t) \sum_{j=1}^n Z_j\, ,
\end{align}
where $Z_i$ are Pauli operators. We are interested in a weak signal limit, where each pulse imprints a small phase $\epsilon  \ll 1$~\footnote{The precise requirements on the size of $\epsilon$ will be clarified below.}, in the presence of local Markovian noise with strength $\gamma$. To illustrate the salient features of our protocol, we consider in the following the case of dephasing noise, such that the Lindblad jump operators $ Z_i$ span the signal Hamiltonian, $H=\mathrm{span}\{ Z_i\}$. That is, the dynamics of the sensors is governed by the Lindblad master equation
\begin{align}
\label{eq:Lindblad}
    \partial_t\rho = -i[H(t),\rho] +\gamma\mathcal{D}[\rho]\ , \ \mathcal{D}[\rho] = \sum_{j=1}^n \big( Z_j\rho Z_j - \rho\big),
\end{align}
where $\mathcal{D}$ represents the local Markovian dephasing noise. Due to the noise, the qubits have a finite coherence time $\tau = 1/\gamma$, which we assume to be much longer than a single pulse duration, $\tau \gg \sigma_t$. For such time scales~$\gg\sigma_t$, the evolution can be modeled with an effective master equation
\begin{align}
\label{eq:master-eff}
\partial_t\bar{\rho} = \gamma\mathcal{D}[\bar{\rho}]+R\,\mathcal{S}_\epsilon[\bar{\rho}]\ 
, \ \mathcal{S}_\epsilon[\bar{\rho}]= U_\epsilon\bar{\rho} U_\epsilon^\dagger - \bar{\rho}
\end{align}
for $\bar{\rho} = \llangle \rho \rrangle$, where $\llangle \cdot \rrangle $ denotes the ensemble average over the Poissonian signal process, and the pulses act as a jump operator $U_\epsilon = \exp(i\epsilon \sum_{i=1}^n Z_i)$ representing collective phase kicks. Note that both signal and noise are therefore effectively described by non-unitary dynamics.

\begin{figure}[t]
    \flushleft
    \includegraphics[width=.95\linewidth]{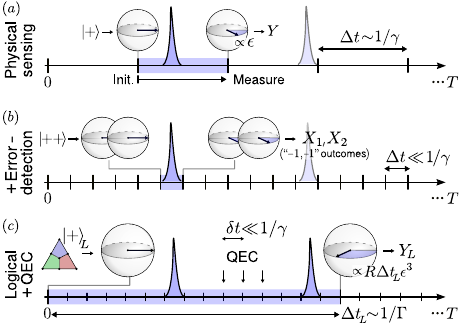}
    \caption{\textit{Sensing strategies.} We compare three different sensing strategies: ($a$)~Physical sensing (Ramsey): Physical quantum sensors are initialized and measured in repeated cycles of duration $\Delta t \sim 1/\gamma$, where $\gamma$ is the physical decoherence rate. ($b$)~Physical sensing with classical error detection: We initialize a pair of physical sensors as $\ket{++}$, forming a classical repetition code. For small exposure time $\Delta t \ll 1/\gamma$, we can exclude (single) physical noise errors by only recording collective qubit flips where $X_1=X_2=-1$. ($c$)~QEC-enhanced logical sensing: We perform syndrome extraction and correction steps in intervals $\delta t \ll 1/\gamma$, to extend the (logical) sensing interval,~$\Delta t_L =1/\Gamma \gg 1/\gamma$, where $\Gamma$ is the logical decoherence rate. During this time, the logical sensor picks up an average phase $\Phi_L \propto R\Delta t_L\epsilon^3$ from multiple, coherently added signal pulses. All three sensing strategies can be enhanced using entanglement between multiple physical/logical sensors.}
    \label{fig:Strategies}
\end{figure}

In the following, we discuss different sensing strategies for this scenario, see~\Fig{fig:Strategies}. We first demonstrate that in the limit $R \ll \gamma $ standard physical Ramsey sensing becomes inefficient, as the rare signals are difficult to distinguish from the noise background. Then, we show how classical error detection and ultimately quantum error correction can help overcome this problem.

\textit{Physical (Ramsey) sensing.---}We first consider a standard Ramsey protocol with physical sensing qubits. For this, we may either use all $n$ qubits independently [case R1], or in an entangled state [case R2].
The qubits are initially prepared in the state $\ket{\psi_{\mathrm{R}1}} = \ket{+}^{\otimes n} \equiv [(\ket{0}+\ket{1})/\sqrt{2}]^{\otimes n}$ [R1], or in the state $\ket{\psi_{\mathrm{R}2}} =\ket{\mathrm{GHZ}_n}\equiv(\ket{0}^{\otimes n}+\ket{1}^{\otimes n})/\sqrt{2}$ [R2]. Subsequently, the state evolves under the Lindbladian \eqref{eq:Lindblad}, is measured in the $Y$-basis after an appropriately chosen sensing interval $\Delta t$ (see below), the data is recorded, and the overall procedure is repeated for a number of $T/\Delta t$ times, see \Fig{fig:Strategies}($a$). 

In the presence of noise, and for $\epsilon \ll 1/n$~\footnote{The sensitivity changes for $\epsilon \sim 1/n$, since, in that regime, sensing intervals with and without signal pulse can be distinguished with single shots with high success probability.}, we choose sensing intervals $\Delta t=1/(2\gamma)$ [R1], and $\Delta t=1/(2\gamma n)$ [R2], which are optimal in this case, and both strategies [R1], [R2] eventually give the same sensitivity~\cite{huelga1997improvement}.
The performance of this protocol to estimate~$\epsilon$ is given by the minimal uncertainty $\sigma_ \epsilon = 1/\sqrt{F_{\mathrm{R}} ^{\mathrm{tot}}}$, where $F_{\mathrm{R}}^{\mathrm{tot}}  = (T/\Delta t) F_\mathrm{R}$, and $F_\mathrm{R}$ is the (quantum) Fisher information of the protocol~\cite{pezze2018quantum}. In total, we get
\begin{align}
\label{eq:classical-F}
    F_\mathrm{R} ^{\mathrm{tot}}  = \frac{nR^2 T}{2e\gamma},\qquad \mathrm{\, [R1]\,\& \,[
R2]}\, ,
\end{align}
with Euler number $e$, see Supplemental Material (SM)~\cite{SM}. The Fisher information is proportional to the number of signal pulses $RT$, and additionally suppressed by the signal-to-noise ratio $R/\gamma \ll 1$. This quantifies the intuitive expectation that rare signals make it increasingly difficult to distinguish signal and noise background.
To eliminate noise, we next consider a classical error-detection protocol.

\textit{Classical error detection protocol.---}We next consider the sensors as a classical error-detecting code, see~\Fig{fig:Strategies}($b$). Again, we distinguish the cases of $n$ physical sensors initialized as unentangled product states $\ket{\psi_{\mathrm{ED}1}} = \ket{+}^{\otimes n}$ [case ED1], and, using entanglement among the sensors, as $\ket{\psi_{\mathrm{ED}2}} = \ket{\mathrm{GHZ}_{n/2}}^{\otimes 2}$ to form two effective, enhanced sensors [case ED2]. The sensors are exposed to both signal and noise for a very short sensing interval $\Delta t \ll 1/(n\gamma)$, after which we measure the individual observables $X_i$, $i=1,..,n$. However, we only record measurement outcomes corresponding to two (or more) $X_i=-1$. This has two effects: First, we filter the noise, as multiple $Z$-errors from dephasing are negligibly rare for $\Delta t\rightarrow 0$. Second, part of the signal is discarded along with the single-sensor decoherence errors. Overall, the probability for recording a signal scales as $p_2\sim [n(n-1)/2]\epsilon^4$ [ED1] and $p_2 \sim [(n/2)\epsilon]^4$~[ED2], which leads to the (classical) Fisher information, see SM~\cite{SM},
\begin{align}
F^{\mathrm{tot}}_{\mathrm{ED}}  &=
\begin{cases}
          8n(n-1) RT\epsilon^2,    &[\mathrm{ED1}],\\[6pt]
 n^4 RT\epsilon^2 ,  &[\mathrm{ED2}]\, .
\end{cases}
\end{align}
We observe that entanglement yields an enhanced sensor performance, i.e. $F_{\mathrm{ED2}}^\mathrm{tot} \geq F_{\mathrm{ED1}}^\mathrm{tot}$. Furthermore, this strategy can outperform the standard physical sensing as $F^\mathrm{tot}_\mathrm{ED2}/F^\mathrm{tot}_\mathrm{R}\sim n^3 \epsilon^2\gamma/R$, if $R$ is sufficiently small. While error detection removes the false positive outcomes due to noise, it however misses signal events, resulting in $F^\mathrm{tot}_\mathrm{ED}\propto\epsilon^2\ll 1$. In the following, we demonstrate how this problem is overcome with a novel QEC-enhanced logical sensing strategy, which allows to extend the logical coherence time $\tau_{L}$ far beyond the physical coherence time of the physical sensors~$\tau$, and which can hence outperform both physical sensing strategies.

\textit{QEC-enhanced logical sensing.---}To illustrate our protocol, we first consider a single logical sensor constructed from seven qubits in a $[[7,1,3]]$ Steane code, and discuss variants with multiple logical sensors later. While the numerical prefactors in the following expressions are specific to the Steane code, the overall scalings are general for distance-3 CSS codes.

\begin{figure}[t!]
    \centering
\includegraphics[width=1.\linewidth]{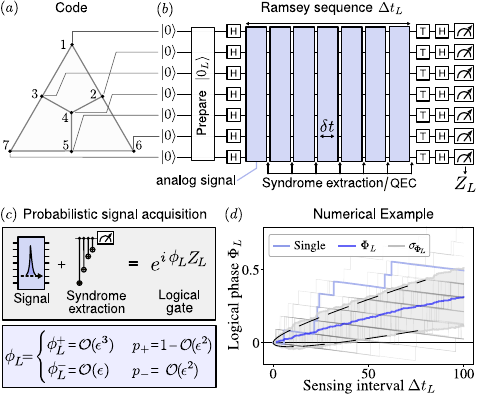}
    \caption{\textit{Logical sensing protocol}. ($a$)~Our quantum sensor is built from a quantum-error-correcting code, e.g. a $[[7,1,3]]$ Steane code. ($b$)~QEC Ramsey cycle: We prepare the logical state~$\ket{+_L}$, then sense the signal for a duration $\Delta t_L$ interspersed with syndrome extraction in intervals of $\delta t$, and, finally, apply a logical projective measurement (choosing the optimal operating/bias point). ($c$)~Syndrome extraction is performed, e.g., by coupling stabilizers to noiseless ancillas, and it converts the signal to a probabilistic logical gate, with logical phase $\phi_L^+/\phi_L^-$ depending on stabilizer outcomes with probabilities $p_+/p_-$. ($d$)~We numerically demonstrate this stochastic acquisition of an average logical signal with phase $ \Phi_L \propto R\Delta t_L \epsilon^3$. Our numerical example shows data for $T=100$, $\delta t=10^{-3}$, $R=30$, $\epsilon= 0.005\times 2\pi$, $\gamma = 0$, and $50$ runs; the dashed lines indicates the analytical estimate~$\Phi_L\pm (R\Delta t_L)\sigma_{\Phi_L}$.}
\label{fig:Scheme}
\end{figure}

Our operation cycle is sketched in \Fig{fig:Strategies}($c$) and \Fig{fig:Scheme}($a$): we perform a Ramsey sequence on the level of a logical qubit over the sensing interval~$\Delta t_L$. That is, we initialize the qubits in the logical state $\ket{\psi_{L}}=\ket{+_L} =(\ket{0_L}+\ket{1_L})/\sqrt{2}$, where $\ket{0_L},\ket{1_L}$ are the logical basis states, and expose all physical sensors to the same signal~\eqref{eq:Hamiltonian}, such that the real-time dynamics on the physical level is again described by~$\mathcal{L}$. During this sensing interval, we perform repeated rounds of syndrome extraction in time steps $\delta t$. The syndrome extraction rate must satisfy $\sigma_t \ll\delta t \ll \tau$ to efficiently correct the dephasing noise; it corrects for single-qubit errors, whereas higher-order errors result in a logical decoherence rate $\Gamma_\gamma= 42\gamma^2 \delta t$.

However, while QEC at such a rate $\sim 1/\delta t $ corrects the noise, it also removes part of the signal. To analyse this, we first consider how the signal interacts with the syndrome extraction rounds in a noise-free setting. That is, we consider an interval $\delta t$ between two syndrome extraction events which contains a signal pulse described by the unitary operation $U_\epsilon=\exp(i\epsilon\sum_i Z_i)$ acting on the qubits. As the Steane code has distance $d=3$, this operator contains a logical signal in third order of $\epsilon$. In the subsequent syndrome measurement, we distinguish two outcomes: With probability $p_+\approx 1-(7/4)\epsilon^2$, all stabilizers are measured with eigenvalue $+1$, and therefore the measurement projects the state back into the code space~$\mathcal{C}$. Importantly, as detailed in the SM~\cite{SM}, this outcome results in the application of a logical phase gate with angle $\phi_L^+ \approx (7/4)\epsilon^3$. With probability $p_-\approx (7/4)\epsilon^2$, however, at least one stabilizer is flipped and the resulting state lies outside of the logical subspace $\mathcal{C}$. We then perform the usual error correction step to bring the state back into $\mathcal{C}$~\footnote{In fact, it is not necessary to perform the error correction step; performing the syndrome measurements suffices.}. As detailed in the SM~\cite{SM}, this operation results in a logical phase gate with angle $\phi_L^- = -3\epsilon$. On average, per signal we thus obtain a logical phase with mean value $\phi_L = p_+ \phi^+_L + p_- \phi^-_L \approx -(7/2)
\epsilon^3$ and variance $\sigma^2_{\phi_L}\approx \tfrac{63}{4}\epsilon^4$. Conversely, for intervals where no pulse appears, the syndrome extraction does not affect the logical state. In total, after a time interval $\Delta t_L$, which contains on average $R\Delta t_L$ signal pulses, the logical sensor acquires the mean logical phase $\Phi_L =(7/2)(R\Delta t_L)\epsilon^3$, see \Fig{fig:Scheme}($c$). The randomness associated with this signal acquisition with stochastic stabilizer outcomes (and their correspondingly different logical phases) results in dephasing of the logical phase given by~$R\Delta t_L\sigma _{\Phi_L}$.

Extending this analysis of the logical sensor to include also the effect of dephasing noise, we find the following logical state after a number of $\Delta t_L/\delta t$ QEC cycles, 
\begin{align}
    \rho_L(\Delta t_L) &\approx \frac{1}{2} \begin{pmatrix}
        1 && e^{(i\Omega-2\Gamma) \Delta t_L} \\ e^{(-i\Omega-2\Gamma) \Delta t_L} && 1
    \end{pmatrix}, 
\end{align}
written in the logical basis $\{\ket{0_L},\ket{1_L}\}$ (see SM~\cite{SM}), and we neglect terms of higher order in $\epsilon$ and $\gamma\delta t$. Here, we defined the logical signal strength $\Omega =\tfrac{7}{2}\epsilon^3 R$, and the logical decoherence rate $\Gamma =\Gamma_\gamma+\Gamma_R$, as the sum of $\Gamma_R= R\sigma^2_{\phi_L}=\tfrac{63}{4} R\epsilon^4$, associated with the stochastic stabilizer outcomes, and $\Gamma_\gamma=42\gamma^2\delta t$ associated with higher-order errors due to the physical noise. Since $\Gamma_\gamma$ scales as $\sim \delta t$, it can be made arbitrarily small compared to $\Gamma_R$ in the limit~$\delta t\rightarrow 0$.
Evaluating the QFI with respect to the phase $\epsilon$ yields
\begin{align}
\label{eq:logical-QFI}
    F_{\mathrm{L}} &= c (R \Delta t_L)^2 \epsilon^4 e^{-2\Gamma \Delta t_L} \, ,
\end{align}
where we have assumed $\gamma\delta t \ll \epsilon^2$, and defined a numerical constant $c=9\times49/4$ specific to the Steane code. Eq.~\eqref{eq:logical-QFI} shows that $F_{\mathrm{L}} \propto \epsilon^4$ which is a simple consequence of $F_{\mathrm{L}}\propto  [\partial\Omega/\partial\epsilon]^2$. This shows that logical sensors behave like single physical sensors but with renormalized signal $\epsilon R\rightarrow \epsilon^3 R$ and dephasing rates $\gamma \rightarrow \gamma^2\delta t$.

\begin{figure}[t!]
    \centering
\includegraphics[width=1.\linewidth]{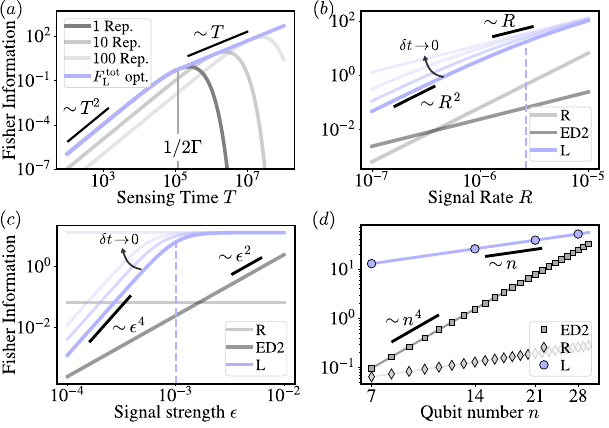}
    \caption{\textit{Sensitivity}. Quantitative comparison of different sensing strategies. ($a$)~Logical QFI $F^\mathrm{tot}_\mathrm{L}$ for $M$ repetitions with $\Delta t_L = T/M$. The blue curve is optimal for given total sensing time $T$ and logical decoherence~$\Gamma$. ($b$)~Sensing performances versus signal rate~$R$. Logical sensing outperforms physical sensing for small $R$ and $\delta t \rightarrow 0$. The dashed vertical line signals the crossover from $R$ to $R^2$ scaling. ($c$)~Similarly, for $\delta t\rightarrow 0$, logical sensing outperforms physical sensing for small~$\epsilon$. The vertical line signals a crossover to $\epsilon$-independence. ($d$)~Error-detection yields the best asymptotic sensitivity scaling with qubit number~$n$. For fixed~$n$, logical sensing outperforms other strategies for sufficiently small $\epsilon$. See SM for numerical parameters~\cite{SM}.}
\label{fig:Protocol}
\end{figure}

From \Eq{eq:logical-QFI} follows, that the optimal logical sensing interval is thus given by $\Delta t_L=1/(2\Gamma)$. Depending on the available total sensing time $T$ we can distinguish two regimes, see \Fig{fig:Protocol}($a$): (\textit{i}) if $T\ll 1/\Gamma$, it is optimal to perform a single sensing step over the full sensing time $\Delta t_L = T$. In this case, the sensitivity is determined via the resulting QFI $F_{\mathrm{L}}^{\mathrm{tot}} = c\,(R  T)^2 \epsilon^4 +\mathcal{O}(\Gamma T)$.
Comparison with the standard physical sensing strategy shows $F^{\mathrm{tot}}_{\mathrm{L}}/F_\mathrm{R}^{\mathrm{tot}}  \sim\epsilon^4\gamma T$. Thus, sensitivity is enhanced by extending the sensing time as $ T \gtrsim 1/(\gamma \epsilon^4)$ beyond the physical coherence time of the sensors. Similarly, comparing against error detection, we obtain $F_\mathrm{L}/F^\mathrm{tot}_\mathrm{ED} \sim RT\epsilon^2$ which can be enhanced for sufficiently large sensing times~$T\gtrsim 1/(R\epsilon^2)$. (\textit{ii}) For even longer evolution times $T\gtrsim 1/\Gamma$, the optimal strategy is to repeat Ramsey cycles of duration $\Delta t_L = 1/(2\Gamma)$ for a number or $T/\Delta t_L$ times, which yields 
\begin{align}
\label{eq:long-time-QFI}
F_{\mathrm{L}}^{\mathrm{tot}}
= c\, \frac{(R\epsilon^2)^2 T}{2\Gamma e} \overset{\delta t\rightarrow 0}{\longrightarrow} c'\frac{RT}{e}\, , 
\end{align}
with $c'=7/2$. Again, this result corresponds to the standard case of noisy individual sensors with corresponding logical signal and dephasing rates. Comparison with the previous sensing protocols shows an improved sensitivity as $F_{\mathrm{L}}^{\mathrm{tot}}/F^{\mathrm{tot}}_\mathrm{R}  \sim \epsilon^4\gamma/\Gamma \rightarrow \gamma/R$, and $F_{\mathrm{L}}^{\mathrm{tot}}/F^{\mathrm{tot}}_\mathrm{ED}  \sim 1/\epsilon^2$, when $\delta t\rightarrow 0$. 
This shows, that the simultaneous renormalization of signal and dephasing rate due to QEC can have a net positive effect in either case: the advantage from the noise renormalization outweighs the (cubic) reduction of the signal. Effectively the underlying reason is that the reduced decoherence yields an extended coherence time $\tau_L = 1/\Gamma \gg \tau$, which allows to extend sensing time as the critical resource for parameter estimation. Note that, in both cases, we are working in the many-pulse regime, where the number of pulses per logical sensing interval is large, i.e., $ RT \gg 1$. Especially in the second case, ($ii$), this number can become so large that the average acquired phase $\Phi_L \sim 1/\epsilon $ exceeds $2\pi$, which needs to be taken into account appropriately.

\textit{Multiple logically entangled sensors.}---Similar to physical sensors, the sensitivity of our protocol can be enhanced by combining multiple logical sensors. If we allow for entanglement between the $n_L$ logical sensors, the optimal strategy and sensing interval $\Delta t_L$ have to be adapted, analogous to the physical sensing case [R2] described above. For this one considers the logical GHZ state, $\ket{\psi_{L}} = (\ket{0_L}^{\otimes n_L}+\ket{1_L}^{\otimes n_L})/\sqrt{2}$, and subsequently applies the sensing protocol to all $n_L$ logical code blocks in parallel. If the total available sensing time is short, $T \ll 1/(n_L\Gamma)$, the optimal sensing interval is again $\Delta t_L = T$, where the extra factor $n_L$ originates from the entanglement-enhanced logical decoherence rate. In the second case, when $T \gtrsim 1/(n_L\Gamma)$, we set the logical sensing interval to $\Delta t_L = 1/(2n_L\Gamma)$. These strategies result in the logical QFI
\begin{align}
F_{\mathrm{L}}^{\mathrm{tot}}  &=
\begin{cases}
          c\, n_L^2 \epsilon^4 (RT)^2,    & T\ll \mathcal{O}[1/(n_L\Gamma)]\, ,\\[6pt]
 c\, n_L(R\epsilon^2)^2 T/(2\Gamma e) ,  &T \ \gtrsim\mathcal{O}[1/(n_L\Gamma)]\, .
\end{cases}
\end{align}
This QFI gives rise to Heisenberg scaling for short evolution times $T$, where the sensitivity improves linearly in $n_L$ and hence linear in the number of physical sensors $n$. At larger times~$T$, this scaling advantage obtained from entangling the~$n_L$ logical sensors is reduced to standard (SQL) scaling in~$n_L$, analogous to the case of physical sensors, where entanglement does not change the asymptotic sensitivity~\cite{huelga1997improvement}. However, there is an advantage of using error-corrected sensors over physical sensing with non-corrected sensors. The sensitivities obtained from the protocols are summarized in~\Fig{fig:Protocol}.

    \textit{Generalizations.---}Our logical quantum sensing scheme can be extended in several ways. First, we can extend our protocol to different error-correcting codes. For instance, we consider a repetition code encoding one logical qubit into an odd number of~$d$ physical qubits. This changes the scaling relations of our protocol. Specifically, as demonstrated in the SM~\cite{SM}, we obtain $F_{\mathrm{L}} \sim (R\Delta t_L)^2 \epsilon^{2(d-1)} \exp(2\Gamma\Delta t_L)$, with the logical decoherence $\Gamma =\Gamma_\gamma+\Gamma_R$, $\Gamma_\gamma \sim \gamma (\gamma\delta t)^{(d-1)/2}$, and $\Gamma_R\sim  R \epsilon^{d+1}$. For $\delta t\rightarrow 0 $, this results in the total QFI, $F_{\mathrm{L}}\sim RT\epsilon^{d-3}$. While this sensitivity is maximized for $d=3$, larger distance can be beneficial for reaching the condition $\Gamma_\gamma/\Gamma_R \ll 1$ at a fixed $\delta t$, as the code can correct for multiple errors. Furthermore, different codes imprint logical phases with different $\epsilon$-dependence, hence, combining codes allows for an independent estimation of $\epsilon$ and $R$, when $R$ is not precisely known.

Second, we can extend our scheme to include more experimentally relevant noise models. For example, this includes quantum noise from spontaneous decay, which is a limiting factor for atomic clocks, as well as Markovian dephasing noise with spatial structure, such as global common-mode noise, see also~\cite{layden2018spatial,layden2019ancilla,sekatski2020optimal,duerDFS,wang2024exponential}\footnote{Such noise processes can also be addressed with decoherence-free subspaces.}.

\textit{Conclusions.---}We have presented a novel sensing protocol which transforms rare and weak transversal rotations into logical gates while preserving quantum coherence with QEC. At first sight, our findings appear to be in conflict with the expectation that QEC cannot enhance sensitivity when signal and noise couple to the sensor with the same operators~\cite{demkowicz2017adaptive,sekatski2017quantum,zhou2018achieving,zhou2021asymptotic}. Here, we explicitly exploit the spatiotemporal structure of the signal: the signal pulses simultaneously and collectively affect all sensor particles. Consequently, the signal generator $U_\epsilon$ can be distinguished from the Lindbladian noise operators in $\mathcal{D}$ through their higher-order dependence on~$\epsilon$. The QEC code performs the required nonlinear data processing that isolates and extracts these higher-order contributions, thereby circumventing the apparent limitations.

While we have demonstrated our protocol for pure dephasing noise, it can be straightforwardly extended to general local Markovian noise with codes that allow to correct for such errors. Note that, in general, our protocol requires to implement active error correction, while for pure dephasing, as shown here, it is sufficient to only extract the syndromes.
A practical realization will require the ability of implementing fast and reliable quantum error correction, where, eventually, the performance will be limited by the pulse duration, the clock-speed of the underlying quantum hardware, and gate-level noise during syndrome extraction.
To conclude, it would be interesting to identify physical use-cases of our protocol to probe rare and weak signals, for instance, in the context of dark matter detection~\cite{jackson2023search} or gravitational wave astronomy~\cite{cahillane2022review}, or, in combination with novel quantum-computing enhanced sensing schemes~\cite{allen2025quantum}.

\textit{Acknowledgments.---}We thank Alexander Baumgärtner, Wolfgang Dür, Charles Fromonteil, Klemens Hammerer, Raphael Kaubrügger, Denis V. Vasilyev, and Peter Zoller for valuable discussions and feedback on the manuscript. This work is supported by the
European Union’s Horizon Europe research and innovation
program under Grant Agreement No. 101113690
(PASQuanS2.1), the ERC Starting grant QARA (Grant
No. 101041435), and the Austrian Science Fund (FWF)
(Grant No. DOI 10.55776/COE1). AMK, HP acknowledge support from the NSF QLCI Award OMA - 2016244, and AMK support from the National Institute of Standards and Technology. SC acknowledges the support from the Sloan Research Fellowship and NSF via QuSeC-TAQS (OMA-2326787) and the Center for Ultracold Atoms, an NSF Physics Frontiers Center (PHY-1734011).

\bibliographystyle{apsrev4-1} 
\bibliography{Bib}

\clearpage

\onecolumngrid

\begin{center}
    \textbf{\large Supplemental Material}
\begin{center}
\begin{minipage}{0.85\textwidth}
This supplemental material provides further information on the (I) the rare-event signal, (II) the standard sensing strategies for this signal, (III) the three-qubit repetition code protocol, (IV) the $[[7,1,3]]$ Steane code protocol, as well as (V) the numerical calculations presented in the manuscript. \\
\end{minipage}
\end{center}

\end{center}
\setcounter{equation}{0}
\renewcommand{\theequation}{S\arabic{equation}}
\setcounter{figure}{0}
\renewcommand{\thefigure}{S\arabic{figure}}
\setcounter{table}{0}
\renewcommand{\thetable}{S\arabic{table}}
\twocolumngrid

\section{Rare-event signal}
\noindent We consider the signal acting on $n$ physical sensing qubits with the Hamiltonian
\begin{align}
    H = \omega(t) \sum_{j=1}^n Z_j\, ,
\end{align}
such that the corresponding time evolution operator $U_\mathrm{sensing}=\mathrm{T}\exp[i\int_0^t \mathrm{d}t' \omega(t') \sum_j Z_j] = \exp[\epsilon N(t)\sum_j Z_j]$ follows a classical Poisson point process with $\llangle N (t) \rrangle =Rt$~\cite{last2018lectures}, where $\llangle \cdot \rrangle$ denotes the ensemble average of the Poisson process, and $\mathrm{T}$ is the time-ordering operator. Additionally, the atoms are subject to Markovian noise with jump operators $L_j=\sqrt{\gamma}Z_j$, and the system evolves with the Lindbladian
\begin{align}
    \mathcal{L} = -i[H,\rho] +\gamma \mathcal{D}[\rho]\ , \quad \mathcal{D}[\rho]= \sum_{j=1}^n \big( Z_j\rho Z_j - \rho\big)\, .
\end{align}
For a time step $\Delta t$, the resulting channel is
\begin{align}
    \rho &\mapsto  \llangle e^{\Delta t \mathcal{L}} \rho \rrangle \nonumber\\
    &= \llangle e^{i\epsilon N(\Delta t) \big[\sum_j Z_j , \ \cdot\,\big]}e^{\Delta t \mathcal{D}} \rho \rrangle \nonumber\\
    &= \llangle e^{i\epsilon N(\Delta t) \big[\sum_j Z_j , \ \cdot\,\big]}\rho_0 \rrangle\, ,
\end{align}
where we used that the unitary and the non-unitary operations commute and we defined $\rho_0=e^{\Delta t \mathcal{D}} \rho$. Using the characteristic function of the Poisson process
\begin{align}
    \llangle e^{i\epsilon N(\Delta t) \mathcal{O}} \rrangle &= e^{R\Delta t[e^{i\epsilon \mathcal{O}}-1]}\nonumber\\
    &\approx (1-R\Delta t)+ (R\Delta t) e^{i\epsilon \mathcal{O}}, 
\end{align}
with an (super-)operator $\mathcal{O}$. For our signal, and using $R\Delta t\ll 1$, we get
\begin{align}
    \rho\mapsto [1-R\Delta t] \rho_0+ [R\Delta t] U\rho_0 U^\dagger\, ,
\end{align}
 with $U(\epsilon) = \exp(i\epsilon \sum_iZ_i)$. This corresponds to the evolution
 \begin{align}
     \dot{\rho} = R\, \mathcal{S}_\epsilon[\rho]= R \big[ U(\epsilon)\rho U^\dagger(\epsilon) - \rho\big]\, .
 \end{align}
 In total, for short $\Delta t$, the system thus evolves with the effective Lindbladian
 \begin{align}
     \mathcal{L}^\mathrm{eff} = R\, \mathcal{S}_\epsilon[\rho] + \gamma\, \mathcal{D}[\rho] \, .
 \end{align}


\section{Standard physical Ramsey sensing}
\noindent In this section, we discuss standard Ramsey quantum sensing protocols in a noisy background. We introduce the overall sensing procedures as well as methods to estimate sensitivities, i.e. the classical and quantum Fisher information. We start by reviewing standard noisy DC sensing, and then discuss complementary strategies for different parameter regimes of our rare-event signal. 

\subsection{DC Ramsey protocol}
\subsubsection{Sensing procedure}
\noindent We first review the standard DC Ramsey protocol, where we assume a signal Hamiltonian $H$ imprinting a phase $\phi$ in a sensing interval $\Delta t$ onto a single-qubit sensor, which we initialize in the state $\ket{\psi_0}=(\ket{0}+\ket{1})/\sqrt{2}$ at the beginning of the sequence. During the sensing interval $\Delta t$, the sensor is subject to both signal and noise, resulting in the mixed state
\begin{align}
    \rho = \frac{1}{2}\begin{pmatrix}
 1  && e^{i\phi - \gamma \Delta t} \\
 e^{-i\phi - \gamma \Delta t} && 1
    \end{pmatrix}\, .
\end{align}
Measurement of $Y$ gives the following outcomes and probabilities
\begin{align}
    Y=+1 &\rightarrow p_+(\phi) = \frac{1-\sin(\phi)e^{-\gamma\Delta t}}{2},\\
    Y=-1 &\rightarrow p_-(\phi) = \frac{1+\sin(\phi)e^{-\gamma\Delta t}}{2}\, .
\end{align}
We define an estimator $\hat{Y}$ for the number of $+1$ outcomes minus the number of $-1$ measurement outcomes, and we obtain
\begin{align}
    \langle \hat{Y}\rangle &=\frac{1-\sin(\phi)e^{-\gamma\Delta t}}{2}-\frac{1+\sin(\phi)e^{-\gamma\Delta t}}{2}\nonumber\\
    &= -\sin(\phi)e^{-\gamma\Delta t}\nonumber\\ &\approx -\phi \,e^{-\gamma\Delta t} \, ,
\end{align}
where we assumed a small phase $\phi\ll\pi$. Inverting this relation yields the unbiased estimator,
\begin{align}
\hat{\phi}=-\hat{Y}e^{-\gamma\Delta t}\, ,
\end{align}
with $\langle\hat{\phi}\rangle = \phi$. Thus, $\phi$ can be straightforwardly extracted from the probabilistic measurement outcomes.

\subsubsection{Sensitivity}
\noindent To estimate the precision of this procedure to extract $\phi$, we first compute the classical Fisher information. To this end, we consider the likelihood $p = p_i(\phi)$ with $i=+,-$, and define the Fisher information as
\begin{align}
    F[p] &= -\sum_i \big[\partial_\phi \log[p_i(\phi)]\big]^2 p_i(\phi)\nonumber\\
    &= -\sum_i \frac{[\partial_\phi p_i(\phi) ]^2}{p_i(\phi)} \, ,
\end{align}
which evaluates to
\begin{align}
    F[p] &= e^{-2\gamma\Delta t}\, .
\end{align}
We compare this result against the quantum Fisher information (QFI) defined as 
\begin{align}
    F[\rho] = 2\sum' _{kl} \frac{\bra{k}\partial_\epsilon \rho \ket{l}\bra{l}\partial_\epsilon \rho \ket{k}}{\lambda_k+\lambda_l}\, ,
\end{align}
where $\lambda_k$ are the eigenvalues of the quantum state $\rho$, and $\sum'$ is restricted to terms which lead to non-zero denominators. Evaluating this expression yields
\begin{align}
    F[\rho] = e^{-2\gamma\Delta t}\, ,
\end{align}
confirming that, previously, we had chosen the optimal measurement basis.

\subsection{Rare-event Ramsey protocol}
\noindent Focusing o our stochastic rare-event signal, we similarly estimate the sensitivity of the Ramsey protocol. We first consider the case of a finite sensing interval $\Delta t$ such that $\epsilon\ll \gamma\Delta t$, and we expand the quantum state in powers of $\epsilon$. In this regime, we compute the quantum Fisher information for a single, as well as multiple unentangled and entangled quantum sensors. Subsequently, we will consider the complementary regime of $\Delta t\rightarrow 0$ with classical error detection further below. 

\subsubsection{Single-qubit sensor}
\noindent For a single-qubit sensor, after a sensing interval~$\Delta t$, our stochastic signal leads to the following mixed state 
\begin{align}
    \rho_\epsilon &=  \tfrac{R\Delta t}{2}\begin{pmatrix}
        1 && e^{-i\epsilon -\gamma \Delta t}\\
        e^{i\epsilon -\gamma \Delta t} && 1
    \end{pmatrix}+  \tfrac{1-R\Delta t}{2}\begin{pmatrix}
        1 && e^{ -\gamma \Delta t}\\
        e^{ -\gamma \Delta t} && 1
    \end{pmatrix}.
\end{align}
For small signals $\epsilon$, we approximate this state as
\begin{align}
    \rho_\epsilon &=  \begin{pmatrix}
        1/2 && e^{ -\gamma \Delta t}\frac{(R\Delta t) (e^{-i\epsilon}-1)+1}{2}\\
        e^{ -\gamma \Delta t} \frac{(R\Delta t) (e^{i\epsilon}-1)+1}{2} && 1/2
    \end{pmatrix} \nonumber\\
    &\approx \begin{pmatrix}
        1/2 && e^{ -\gamma \Delta t}\frac{ 1 - i\epsilon(R\Delta t)}{2}\\
        e^{ -\gamma \Delta t} \frac{ 1 + i\epsilon(R\Delta t)}{2} && 1/2
\end{pmatrix}\nonumber\\
    &\approx \frac{1}{2}\begin{pmatrix}
        1 && e^{ -\gamma \Delta t-i\epsilon(R\Delta t)} \\
        e^{ -\gamma \Delta t +i\epsilon(R\Delta t)} && 1
    \end{pmatrix} .
\end{align}
The QFI of this state evaluates to
\begin{align}
    F_\mathrm{R}[\rho_\epsilon] = (R\Delta t)^2 e^{-2\gamma\Delta t}\, .
\end{align}
For a total sensing time $T$, we repeat this protocol $T\Delta t$ times, which yields
\begin{align}
    F_\mathrm{R}^{\mathrm{tot}}[\rho_\epsilon] = \frac{T}{\Delta t} R^2 (\Delta t)^2 e^{-2\gamma \Delta t}\, .
\end{align}
This expression is maximized for
\begin{align}
    \frac{\partial F_\mathrm{R}^{\mathrm{tot}}}{\partial (\Delta t)} &= R^2T e^{-2\gamma\Delta t}\left[ -2\gamma(\Delta t) + 1\right] \overset{!}{=} 0, \\
    \rightarrow \Delta t &= \frac{1}{2\gamma} .
\end{align}
Therefore, within the discussed approximations, the QFI at the optimal interrogation time is
\begin{align}
F_\mathrm{R}^{\mathrm{tot}} = \frac{R^2 T}{2\gamma e}.
\end{align}

\subsubsection{Multiple qubits}
\noindent We next consider the above scenario with $n$ identical physical sensors. We first consider the case of using $n$ independent (unentangled) sensors in parallel. After the Ramsey cycle, the corresponding state reads
\begin{align}
    \rho^{[n]}_\epsilon \approx (R\Delta t) \rho_\epsilon^{\otimes n} + (1-R\Delta t)\rho_0^{\otimes n}\, ,
\end{align}
where $\rho_\epsilon,\rho_0$ are the single-qubit states with and without applied signal $U(\epsilon)$ respectively. To compute the QFI, we approximate the full state as
\begin{align}
    \rho^{[n]}_\epsilon = \rho_0^{[n]}+ \epsilon (\partial_\epsilon \rho^{[n]}_\epsilon)\big|_{\epsilon=0} + \mathcal{O}(\epsilon^2)\, .
\end{align}
We use the product rule
\begin{align}
    \partial_\epsilon \rho^{[n]}_\epsilon\big|_{\epsilon=0} &= \tfrac{R\Delta t}{2} \Big[ \begin{pmatrix}
        0 && -i e^{-\gamma\Delta t} \\ i e^{-\gamma\Delta t} && 0
    \end{pmatrix}\otimes \rho_0 \otimes \ldots + \ldots\Big] \nonumber\\
    &= \tfrac{R\Delta t }{2}  e^{-\gamma \Delta t}\big[ \sigma_y \rho_0^{\otimes (n-1)} + \rho_0\,  \sigma_y \, \rho_0^{\otimes (n-2)}+ .. \big] .
\end{align}
Using this, the state is approximated as
\begin{align}
    \rho_\epsilon^{[n]} &= \rho_0^{\otimes n} + \tfrac{R\Delta t \epsilon e^{-\gamma \Delta t}}{2}\big[ \sigma_y \rho_0^{\otimes (n-1)} + \rho_0 \otimes \sigma_y\otimes \rho_0^{\otimes (n-2)}+ .. \big]\nonumber\\
    &= \rho^{\otimes n}_{(R\Delta t \epsilon)} +\mathcal{O}(\epsilon^2)\, ,
\end{align}
where we have used the relation
\begin{align}
    \rho_{R\Delta t\epsilon} = \rho_0 + \tfrac{R\Delta t\epsilon}{2} e^{-\gamma \Delta t} \sigma_y +\mathcal{O}[(R\Delta t\epsilon)^2]\, .
\end{align}
Note that, in this derivation, we have neglected terms of order $\mathcal{O}(\epsilon^2)\times \mathcal{O}(n^2)$, and it is thus expected to be accurate for $\epsilon \ll 1/n$. Overall, within these approximations, the QFI is given by
\begin{align}
    F_\mathrm{R} =  n (R\Delta t)^2 e^{-2\gamma\Delta t}\quad [\mathrm{R}1]\, ,
\end{align}
in the case of unentangled sensors [R1].\\

\noindent 
Similarly, in case [R2], we may use the $n$ atoms to form an entanglement-enhanced quantum sensor by initializing the atoms in a GHZ state $\ket{\psi_0} = (\ket{0}^{\otimes n}+\ket{1}^{\otimes n})/\sqrt{2}$. This is identical to the single-qubit case with both enhanced phase $\phi\rightarrow n\phi$ and noise rate $\gamma\rightarrow n\gamma$. Therefore, for a GHZ sensor one obtains
\begin{align}
     F_\mathrm{R} = (nR\Delta t)^2 e^{-2n\gamma\Delta t}\quad [\mathrm{R}2]\, .
\end{align}
Optimizing the sensing time for multiple repetitions within a fixed sensing time $T$, yields the expressions stated in the main text for both cases,
\begin{align}
\label{eq:classical-F}
    F_\mathrm{R}^{\mathrm{tot}}  = \frac{nR^2T}{2\gamma e}\qquad \mathrm{[R 1\,\&\, R2]}\, .
\end{align}

\subsection{Classical error detection}
\noindent In this section we discuss a sensing protocol based on a classical repetition code and measurement post-selection. In contrast to previous sections, we address the regime of infinitesimally short $\Delta t$, where our previous expansion of the quantum state in powers of $\epsilon$ fails to remain valid.

\subsubsection{Two physical sensors}
\noindent We consider a setup with two identical physical sensors. We first compute the classical Fisher information for a classical error detection procedure. That is, we prepare both physical sensors in the state $\ket{+}$, expose the sensors to both signal and noise, and measure the observables $X_1$, $X_2$, after a sensing interval $\Delta t$. In the limit $\Delta t\rightarrow 0$ we neglect the possibility of having multiple noise errors, resulting in the approximate measurement probabilities
\begin{align}
    &(X_1,X_2)=(1,1) &&\quad p_{11} = 1 - \mathcal{O}(\Delta t)\, , \\
    &(X_1,X_2)=(1,-1) &&\quad p_{1-1} = \epsilon^2\, R\Delta t + \gamma\Delta t\, ,\\    
    &(X_1,X_2)=(-1,1) &&\quad p_{-11} = \epsilon^2\,R\Delta t + \gamma\Delta t\, ,\\    
    &(X_1,X_2)=(-1,-1) &&\quad p_{-1-1} = \epsilon^4\,R\Delta t + \mathcal{O}(\Delta t^2)\, .
\end{align}
In the following, we consider the likelihood function $\mathrm{p}$ for the two events `-1-1' and `Not -1-1', with $\mathrm{p}_0(\epsilon) = p_{-1-1}$ and $\mathrm{p}_1(\epsilon)=1-p_{-1-1}$. This corresponds to classically discard all `single-sensor-flip' events which can be interpreted as classical error detection. 
The associated classical Fisher information evaluates to
\begin{align}
    F[\mathrm{p}] &= - \sum_{i=0}^1 \frac{[\partial_\epsilon \mathrm{p}_i(\epsilon)]^2}{\mathrm{p}_i(\epsilon)}\nonumber\\
    &\approx \frac{[4\epsilon^3 R\Delta t]^2}{\epsilon^4R\Delta t} + \frac{[4\epsilon R\Delta t]^2}{1}\nonumber\\
    &\approx 16 \epsilon^2 R\Delta t \, .
\end{align}

\subsubsection{Multiple physical sensors}
\noindent Repeating the same procedure with $n$ identical unentangled physical sensors [ED1], yields the following measurement outcomes and probabilities
\begin{align}
    &\mathrm{all}\ X_i\ = \ 1 &&\quad p_0 = 1 - \mathcal{O}(\Delta t)\\
    &\mathrm{one}\ X_i= -1 &&\quad p_1 = n [\epsilon^2 R\Delta t +  \gamma\Delta t]\\
    &\mathrm{two}\ X_i= -1 &&\quad p_2 = \tfrac{n(n-1)}{2} [\epsilon^2 R\Delta t +  \gamma\Delta t]^2\,  
\end{align}
The corresponding classical Fisher information for the event of multiple $X_i=-1$ outcomes results in
\begin{align}
    F_{\mathrm{ED}} = 8n(n-1) \epsilon^2 R\Delta t \ \quad [\mathrm{ED1}]\, .
\end{align}
Next, we consider the case of entangled sensors [ED2]. Suppose we split the $n$ physical sensors into two collective sensors of $n/2$ physical sensors each. We initialize each of these collective sensors as $\ket{\mathrm{GHZ}_{n/2}}$. As a result, we obtain the same result as the two-sensor case, including the replacement $\epsilon \rightarrow (n/2)\epsilon$ and $\gamma\rightarrow (n/2)\gamma$. That results in the classical Fisher information 
\begin{align}
F_{\mathrm{ED}}=n^4\epsilon^2R\Delta t \ \quad [\mathrm{ED2}]\, .
\end{align}
\section{Details on the Repetition code}

\noindent In this section, we analytically derive the sensitivity and logical decoherence rate of the QEC Ramsey protocol. For simplicity, in this section, we focus on a three-qubit repetition code, which protects against $Z$-errors. That is, the code words are $\{\ket{+++},\ket{---}\}$, and we have three X-stabilizers $S_1=X_1X_2$, $S_2=X_2X_3$, and $S_2=X_1X_3$, as well as a logical operator $X_L=X_1X_2X_3$. The code words span the two-dimensional logical code space $\mathcal{C}$ in which we will record the signal as a logical relative phase. We first discuss the signal acquisition in this code, then the error correction, and, at last, we compute the quantum Fisher information for this case.

\subsection{Logical phase gate from weak transversal rotations}
\noindent We first discuss how the physical signal, acting as a weak transversal rotation, gets promoted to a logical phase gate. To this end, we first consider the application of the signal $U(\epsilon)$ followed by stabilizer measurements and error correction procedures and we consider the case without noise first.\\
\noindent If we perform a stabilizer measurement after a pulse signal, there are four different possible outcomes $(i,j)$ with $i,j\in\{1,-1\}$, and the measurement projects into the corresponding subspace $\mathcal{C}_{ij} = P_{ij}\mathcal{H}$ with projectors $P_{ij}$ and probabilities $p_{ij}$, where $\mathcal{C}_{11}=\mathcal{C}$ corresponds to the code space. For simplicity, here we assume that after each round of measurement, we perform an error correction step $E_{ij}$ corresponding to the stabilizer outcome ($i,j$), thus bringing the quantum state back into $\mathcal{C}$. In practice, these correction steps are not necessary, since we can classically adapt the Pauli frame after each measurement step.

\noindent To describe the action of the signal on the code space, we first expand the signal $U$ as
\begin{align}
    U(\epsilon) &= \prod_{i=1}^3[\cos(\epsilon)+i\sin(\epsilon)Z_i]\\
    &= \cos^3(\epsilon)+i^3\sin^3(\epsilon) Z_L + i\sin(\epsilon) \cos(\epsilon)\big(\sum_{i=1}^3 Z_i \big)e^{i\epsilon Z_L} \nonumber
\end{align}
To see how the unitary $U$ interplays with the stabilizer measurements, we decompose it into the different stabilizer-subspaces as
\begin{align}
    (1,1):\, P_{11}UP_{11} &=  \cos^3(\epsilon)+i^3\sin^3(\epsilon) Z_L\nonumber\\& \approx (1-\tfrac{3}{2}\epsilon^2) e^{-i\epsilon^3 Z_L} ,\\
    (1,-1): P_{1-1}UP_{11} &= Z_3 i\sin(\epsilon) \cos(\epsilon)e^{i\epsilon Z_L}\nonumber\\
    \rightarrow E_{1-1}P_{1-1}UP_{11}&= i\sin(\epsilon) \cos(\epsilon)e^{i\epsilon Z_L} \, .
\end{align}
After detecting the outcome $(1,-1)$, corresponding to the error $Z_3$, it is removed by applying the error correction step $E_{1-1}$ in the form of an additional $Z_3$ operator. Analogously, for the other two cases $(-1,1)$ and $(-1,-1)$ we obtain
\begin{align}
    (-1,1): E_{-11}P_{-11}UP_{11} &=  i\sin(\epsilon) \cos(\epsilon)e^{i\epsilon Z_L},\nonumber\\
    (-1,-1): E_{-1-1}P_{-1-1}UP_{11} &= i\sin(\epsilon) \cos(\epsilon)e^{i\epsilon Z_L}\, ,
\end{align}
and, again, the single-qubit errors detected in the syndrome measurements are removed by the corresponding error-correction operation.
To leading order, this implies the following combination of probabilities and resulting logical phases for the evolution of the logical state $\rho_L$,
\begin{align}
    (1,1):& \quad p_{11}= 1-3\epsilon^2\quad &&\phi^+_L  = -\epsilon^3, \nonumber\\
    (1,-1):& \quad p_{1-1}= \epsilon^2\quad &&\phi^-_L = \epsilon ,\nonumber\\
    (-1,1):& \quad p_{-11}= \epsilon^2\quad &&\phi^-_L = \epsilon, \nonumber\\
    (-1,-1):& \quad p_{-1-1}= \epsilon^2\quad &&\phi^-_L = \epsilon\, .
\end{align}
On average, to leading non-vanishing order in $\epsilon$, we expect to pick up the following logical phase for each signal event
\begin{align}
  \phi_L &= p_{+}\phi^+_L + p_-\phi^-_L \nonumber\\
  &= 2\epsilon^3 + \mathcal{O}(\epsilon^5)\, ,
\end{align} 
where $p_+=p_{11}$ is the probability for no error, and $p_{-} = p_{1-1}+p_{-11}+p_{-1-1}$ is the probability for detecting the stabilizers corresponding to one error.
If we apply this operation (i.e., $U$ followed by syndrome extraction and correction) to a density operator in the logical subspace $\mathcal{C}$, i.e. $\rho_L = P_{11}\rho_L = \rho_LP_{11}$, we obtain the following channel,
\begin{align}
    \rho_L &\mapsto \mathcal{U}[\rho_L]=
    \sum_{ij}  E_{ij}P_{ij} U\rho_L U^\dagger P_{ij}E_{ij} 
    \\
    &=[1-3\epsilon^2] e^{-i\phi^+_LZ_L}\rho_L e^{i\phi^+_LZ_L} + [3\epsilon^2] e^{-i\phi^-_LZ_L}\rho_L e^{i\phi^-_LZ_L}\nonumber \, .
\end{align}
Expanding this expression, one can show that this is equivalent to a logical unitary operation $U_{L}$ in combination with an effective logical dephasing $\mathcal{D}_L$, given by
\begin{align}
    \mathcal{U}[\rho_L] \approx  \mathcal{E}_{L1}
[U_L\rho_LU_L^\dagger], \quad U_L = e^{-i \phi_L Z_L},\quad  \mathcal{E}_{L1} = e^{3\epsilon^4 \mathcal{D}_L},
\end{align}
where we defined $\mathcal{D}_L[\rho_L] = Z_L\rho_LZ_L -\rho_L$.\\

\noindent In our sensing scheme, the signal $U(\epsilon)$ is rare and arrives with small rate $R$. That is, the signal channel, as defined previously, corresponds to a Lindbladian
\begin{align}
    R\,\mathcal{S}_\epsilon[\rho_L] = R[U(\epsilon)\rho_L U^\dagger(\epsilon) -\rho_L] \, .
\end{align}
It can be straightforwardly shown, that, taking into account the syndrome extraction and error correction steps outlined in this section, the logical state evolves according to the Lindbladian 
\begin{align}
    R\,\mathcal{S}_L[\rho_L] = R\,\mathcal{S}_{2\epsilon^3}[\rho_L] + 3\epsilon^4R\,\mathcal{D}_L[\rho_L]
\end{align}
where we took into account the leading order (in $\epsilon$) signal and dephasing processes.


\subsection{Noise}
\noindent In this section, we consider the interplay of the three-qubit repetition code with the physical decoherence noise. Specifically, the code allows us to detect and correct single-qubit errors, while higher-order error processes lead to logical decoherence. In the following, we will derive this effective logical decoherence rate.

\noindent Solving the Lindbladian evolution $\exp(\delta t \gamma \mathcal{D})$ in the regime $\gamma \delta t \ll 1$ for a three-qubit repetition code yields
\begin{align}
    \rho_L \mapsto &[1-3\gamma\delta t - 6(\gamma\delta t)^2]\rho_L\nonumber\\ &+ (\gamma\delta t) \sum_{i=1}^3 Z_i\rho_L Z_i
    + (\gamma\delta t)^2 \sum_{i\neq j=1}^3 Z_jZ_i\rho_L Z_iZ_j \, ,
\end{align}
where we neglect higher-order errors $\sim \mathcal{O}[(\gamma\delta t)^3]$. This expression corresponds to the following error probabilites
\begin{align}
    0\ \mathrm{errors}\quad &p_0= 1 - 3(\gamma\delta t) -6(\gamma\delta t)^2 \nonumber\\
    1\ \mathrm{error\ }\quad &p_1=3(\gamma\delta t) \nonumber\\
    2\ \mathrm{errors}\quad &p_2=6(\gamma\delta t)^2\, ,
\end{align}
and these numbers include the multiplicity of having errors in different physical qubits, such that $\sum_{i=0}^2 p_i = 1$. Quantum error correction detects errors by measurements which project the quantum state on subspaces $\mathcal{C}_{ij}$. It is able to detect and correct the single-qubit errors, however, two-qubit errors are promoted to logical errors. Hence, at the logical level we have
\begin{align}
    0\ \mathrm{logical\ errors}\quad &p_{L0}= 1 -6(\gamma\delta t)^2 \nonumber\\
    1\ \mathrm{logical\ error\ }\quad &p_{L1}=6(\gamma\delta t)^2 \, .
\end{align}
Therefore, the resulting channel in this step is given by
\begin{align}
    \rho_L \mapsto \mathcal{E}_{L2}[\rho_L] &=
    \sum_{ij} E_{ij} P_{ij} e^{\delta t \gamma \mathcal{D}}[\rho_L]  P_{ij}E_{ij} 
    \\
    &\approx [1 -6(\gamma\delta t)^2]\rho_L + [6(\gamma\delta t)^2] Z_L \rho_L Z_L\nonumber \, .
\end{align}
This is equivalent to a logical dephasing channel $\mathcal{E}_{L2} = e^{\Gamma_\gamma\delta t \mathcal{D}_L}$, with $\Gamma_\gamma =6 \gamma^2\delta t$ within one error-correction interval $\delta t$, and, similarly, we obtain $\mathcal{E}_{L2} = e^{\Gamma_\gamma\Delta t_L \mathcal{D}_L}$ for a longer, logical sensing interval $\Delta t_L$.

\noindent\textit{Note:} In all of these cases we have written the state after measurement as an ensemble without including the information of the measurement outcome. In fact, we do use the information of the measurement outcome to perform error correction, but otherwise we do not get substantial information from these measurements since the errors are mostly originating from the dephasing noise.

\subsection{Combining signal and noise}
\noindent In the next step, we combine the action of signal and dephasing errors, followed by syndrome extraction and error correction. As both the signal and the errors from dephasing
are rare, the outcome is that, to leading order, both effects add up independently. That is, for a given error-correction interval $\delta t$ the probability of having a signal $R\delta t$, and the probability of having a dephasing error $\gamma \delta t$, are individually much larger than the probability of having both simultaneously, $R\gamma (\delta t)^2$. 

\noindent Our protocol combines both effects in a single channel with Lindbladian $\Lambda$,
\begin{align}
    \rho_L &\mapsto e^{(R\Delta t_L)\Lambda}[\rho_L]\, ,\\
    \Lambda[\rho_L] &\approx R\, \mathcal{S}_{2\epsilon^3}[\rho_L]+ (6\gamma^2\delta t + 3\epsilon^4 R)\mathcal{D}_L[\rho_L]
\, .
\end{align}

\subsection{Logical QFI}
\noindent To assess the sensitivity of our protocol, we compute the QFI associated with the logical quantum state after the action of the Lindbladian $\Lambda$. To this end, we initialize the quantum state of a single logical sensor in the logical subspace as $\rho_L(0)=\ket{+_L}\bra{+_L}$, which, written in the logical basis $\{\ket{0_L},\ket{1_L}\}$, reads
\begin{align}
    \rho_L(0) &\approx \frac{1}{2} \begin{pmatrix}
        1 && 1\\ 1 && 1
    \end{pmatrix}.
\end{align}
It can be easily shown, that the application of the Lindbladian $\Lambda$ leads to the logical quantum state, given by
\begin{align}
    \rho_L(\Delta t_L) &= \frac{1}{2} \begin{pmatrix}
        1 && e^{(i\Omega-2\Gamma) \Delta t_L} \\ e^{(-i\Omega-2\Gamma) \Delta t_L} && 1
    \end{pmatrix}, 
\end{align}
as stated in the main text, where, in leading non-vanishing order, $\Omega = 2\epsilon^3R$, and $\Gamma = 6\gamma^2 \delta t + 3\epsilon^4 R$ for the three-qubit repetition code. This state is analogous to the quantum state of a physical sensor in a DC-sensing protocol, with renormalized frequency $\Omega$ and decoherence rate $\Gamma$. As such, the QFI with respect to $\epsilon$ is given by
\begin{align}
    F_L &= \Big[\frac{\partial (\Omega\Delta t_L)}{\partial\epsilon}\Big]^2e^{-2\Gamma\Delta t_L}\nonumber\\
    &=36\epsilon^4 R^2 \Delta t_L^2 e^{-2\Gamma\Delta t_L}\, .
\end{align}

\section{Details on the Steane code}
\noindent In this section we describe the logical sensing protocol for the $[[7,1,3]]$ Steane code. It encodes one logical qubit into seven physical qubits and has a code distance 3, that is, it can correct for one single-qubit error. Pictorially, the Steane code is given by
\begin{align}
\mathrm{Steane} = \raisebox{-10pt}{\includegraphics[width=.15\columnwidth]{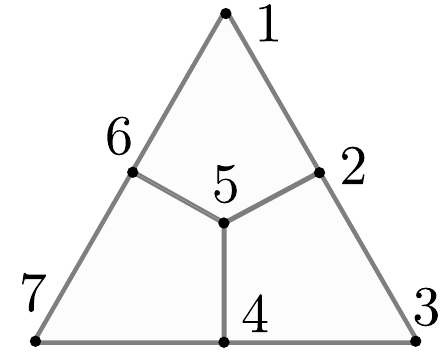}}
\end{align}
The Steane code has two logical code words,
\begin{align}
    \ket{0_L} &= \frac{1}{\sqrt{8}} \big[ \ket{0000000}+\ket{1010101}+\ket{0110011}+\ket{1100110} \nonumber\\
&\quad +\ket{0001111}+\ket{1011010}+\ket{0111100}+\ket{1101001} \big]\nonumber\, ,\\
    \ket{1_L} &= \frac{1}{\sqrt{8}} \big[ \ket{1111111}+\ket{0101010}+\ket{1001100}+\ket{0011001}\nonumber\\
&\quad +\ket{1110000}+\ket{0100101}+\ket{1000011}+\ket{0010110} \big]\, ,
\end{align}
stabilized by the operators $S^X_{1}=X_1X_2X_5X_6$, $S^X_{2}=X_2X_3X_4X_5$, and $S^X_{3}=X_4X_5X_6X_7$. Pictorially, we can illustrate these states as
\begin{align}
    \ket{0_L} &= \frac{1}{\sqrt{8}} \Big[ \, \raisebox{-8pt}{\includegraphics[width=.55\columnwidth]{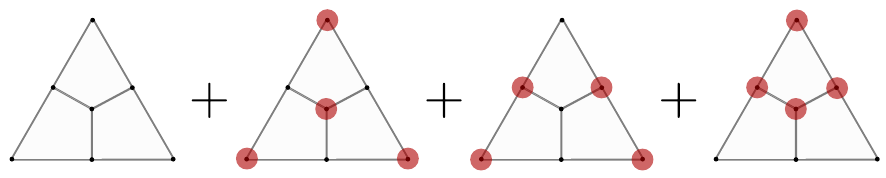}} \nonumber\\
&\quad +\raisebox{-8pt}{\includegraphics[width=.55\columnwidth]{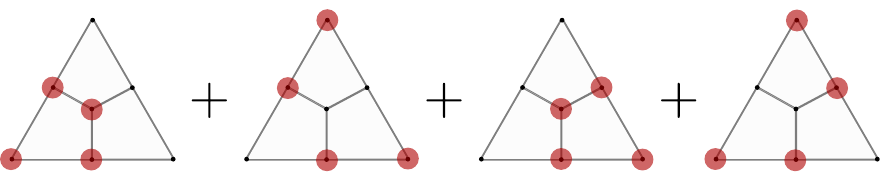}} \,\Big]\, ,\\
    \ket{1_L} &= \frac{1}{\sqrt{8}} \Big[ \, \raisebox{-8pt}{\includegraphics[width=.55\columnwidth]{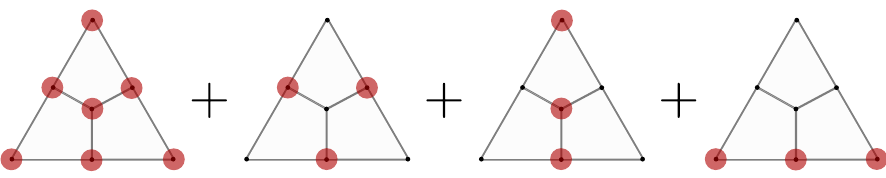}} \nonumber\\
&\quad +\raisebox{-8pt}{\includegraphics[width=.55\columnwidth]{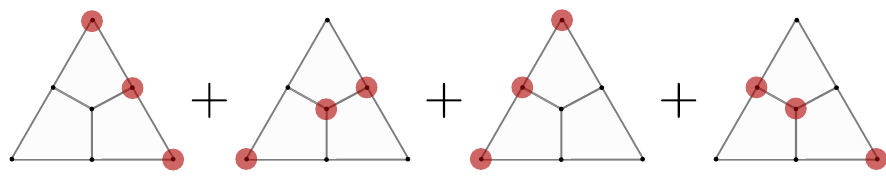}} \,\Big]\, ,
\end{align}
where red circles correspond to qubits in state $\ket{1}$, otherwise, qubits are in state $\ket{0}$.

\subsection{Apply signal}
\noindent To assess the action of the signal in combination with the syndrome extraction and error-correction steps, we first apply the signal $U(\epsilon) =\exp(i\epsilon \sum_{i=1}^7 Z_i)$ to the logical basis states. We obtain
\begin{align}
    &U\ket{0_L} \nonumber\\
    &= \frac{e^{-i\epsilon}}{\sqrt{8}} \big[ e^{i8\epsilon}\ket{0000000}+\ket{1010101}+\ket{0110011}+\ket{1100110} \nonumber\\
&\quad +\ket{0001111}+\ket{1011010}+\ket{0111100}+\ket{1101001} \big]\nonumber\, ,
\end{align}
\begin{align}
    &U\ket{1_L}\nonumber\\
    &= \frac{e^{i\epsilon}}{\sqrt{8}} \big[e^{-i8\epsilon} \ket{1111111}+\ket{0101010}+\ket{1001100}+\ket{0011001}\nonumber\\
&\quad +\ket{1110000}+\ket{0100101}+\ket{1000011}+\ket{0010110} \big]\, .
\end{align}
From this, we see clearly that $\epsilon_T=\pi/4$ represents a transversal logical $S$-gate, $U(\pi/4) = S_L$. In the following, we are interested in the limit $\epsilon \ll \epsilon_T$.

\subsection{Stabilizer measurement}
\noindent As a next step, we compute the possible outcomes of a stabilizer measurement following the application of the signal $U(\epsilon)$. In analogy to the repetition code example, we define the projectors $P_{ijk}$ onto the subspace $\mathcal{C}_{ijk}$, where $i,j,k=\{-1,1\}$ correspond to the eigenvalues of the stabilizers $S^{X}_1$, $S^{X}_2$, and $S^{X}_3$. We obtain,
\begin{align}
        P_{111}U\ket{0_L} &= [\frac{7
        +e^{i4\epsilon}}{8}] \ket{0_L}\, ,\\ P_{111}U \ket{1_{L}}&= e^{i\epsilon}[\frac{7
        +e^{-i4\epsilon}}{8}] \ket{1_L}\, .
    \end{align}
The probability for the corresponding measurement outcome $(1,1,1)$ is given by
\begin{align}
        p_+ =p_{111} &= \Big| \frac{7+e^{\pm i4\epsilon}}{8}\Big|^2  \approx 1 - \frac{ 7}{4}\epsilon^2  +\mathcal{O}(\epsilon^3)\, ,
\end{align}
and the procedure implements the logical phase
    \begin{align}
        \phi_L^+ &=\arg\!\left(e^{i\epsilon}\frac{7+e^{-i4\epsilon}}{7+e^{i4\epsilon}}\right)= \frac{7}{4}\epsilon^3 +\mathcal{O}(\epsilon^5) \;\, .
     \end{align} 
Similarly, projecting onto a different stabilizer outcome, e.g. $(1,1,-1)$, yields, 
\begin{align}
        P_{11-1}\ket{0_L} &=   [\frac{1-
        e^{i4\epsilon}}{8}]\ket{0_L}\, ,\\ 
        P_{11-1}\ket{1_L} &=   -Z_7e^{i\epsilon}[\frac{1-
        e^{-i4\epsilon}}{8}]\ket{1_L}\, ,
    \end{align}
and subsequent error correction (with $E_{11-1}$) removes the $Z_7$ operator. The probability for this measurement outcome is
\begin{align}
        p_{11-1} &= |\frac{1-e^{i4\epsilon}}{8}|^2\approx  \frac{\epsilon^2}{4} +\mathcal{O}(\epsilon^3) \, ,
\end{align}
and the logical phase given by
\begin{align}
        \phi_L^- &=\arg\!\left(-e^{i\epsilon}\frac{1-e^{-i4\epsilon}}{1-e^{i4\epsilon}}\right)=-3\epsilon \;\, .
\end{align} 
All other stabilizer outcomes are analogous, such that the probability of projecting into a subspace orthogonal to $\mathcal{C}$ is given by
\begin{align}
    p_- = 7 p_{11-1} \approx \frac{7}{4} \epsilon^2\, .
\end{align}
Using these resulting phases and probabilities, we obtain on average
\begin{align}
    \phi_L &= p_+ \phi_L^+ + p_- \phi_L^- \approx -\tfrac{7}{2} \epsilon^3 \, ,\\
    \sigma^2_{\phi_L} &= p_+ (\phi^+_L - \phi_L)^2 + p_- (\phi^-_L - \phi_L)^2 \approx \tfrac{63}{4} \epsilon^4\, .
\end{align}

\subsection{Noise}
\noindent Similar to the case of the three-qubit repetition code, we obtain a logical dephasing rate from higher-order error processes. Again, solving the Lindbladian evolution $\exp(\delta t \gamma \mathcal{D})$ in the regime $\gamma \delta t \ll 1$ for the seven physical sensors of the Steane code, yields
\begin{align}
    \rho_L \mapsto &[1-7\gamma\delta t - 42(\gamma\delta t)^2]\rho_L\nonumber\\ &+ (\gamma\delta t) \sum_{i=1}^7 Z_i\rho_L Z_i
    + (\gamma\delta t)^2 \sum_{i\neq j=1}^7 Z_jZ_i\rho_L Z_iZ_j \, ,
\end{align}
and we again neglect higher-order errors $\sim \mathcal{O}[(\gamma\delta t)^3]$. As before, the error correction procedure yields the effective evolution
\begin{align}
    \rho_L \mapsto \mathcal{E}_{L2}[\rho_L] &=
    \sum_{ij} E_{ij} P_{ij} e^{\delta t \gamma \mathcal{D}}[\rho_L]  P_{ij}E_{ij} 
    \\
    &\approx [1 -42(\gamma\delta t)^2]\rho_L + [42(\gamma\delta t)^2] Z_L \rho_L Z_L\nonumber \, .
\end{align}

\subsection{Logical QFI}
\noindent To summarize, we hence obtain the logical frequency $\Omega = -\tfrac{7}{2}\epsilon^3 R$, and logical decoherence rate $\Gamma = 42\gamma^2 \delta t + \tfrac{63}{4}\epsilon^4 R$ for the Steane code. The logical QFI resulting from these expressions is given by 
\begin{align}
    F_\mathrm{L} & =\tfrac{21^2}{4}\epsilon^4 R^2 \Delta t_L^2 e^{-2\Gamma\Delta t_L}\, .
\end{align}

\section{Numerical calculations}
\noindent In this section we briefly describe the numerical data shown in Fig.~2 of the main text, as well as the parameters used for the curves shown in Fig.~4 of the main text.

\subsubsection*{Fig.~2}
\noindent In this figure, we illustrate the coherent but stochastic phase acquisition in the Steane code. To this end, we simulate the real-time dynamics of the 7 qubits with exact diagonalization. 

\noindent We chose the parameters: 
\noindent $\Delta t_L=100$, $\delta t = 10^{-3}$, $R=30$, $\epsilon = 5\times 10^{-3}$, $\gamma = 0$.
\subsubsection*{Fig.~4}
\noindent In Fig.~4 we plot the analytical expressions as given in main text. To this end, we use a combination of physical parameters which illustrate the salient features of the different protocols.\\

\noindent The chosen parameters are:

\noindent Panel ($a$): We choose the parameters: $\gamma = 10^{-2}$, $\epsilon = 10^{-2}$, $n = 7$, $R=10^{-2}$, and $\delta t = 10^{-3}$.

\noindent Panel ($b$): We choose the parameters: $\gamma = 2\times 10^{-4}$, $\epsilon = 10^{-3}$, $n = 7$, $T=10^7$, and $\delta t = 2.5\times 10^{-11}$.

\noindent Panel ($c$): We choose the parameters: $\gamma = 2\times 10^{-4}$, $n = 7$, $T=10^7$, and $\delta t = 10^{-11}$, $R=10^{-6}$.

\noindent Panel ($d$): We choose the parameters: $\gamma = 2\times 10^{-4}$, $T=10^7$, and $\delta t = 0$, $R=10^{-6}$, and $\epsilon = 2\times 10^{-3}$.

\end{document}